\documentstyle[12pt]{article}
\def\be{\begin{eqnarray}}
\def\ee{\end{eqnarray}}
\def\ba{\begin{array}}
\def\ea{\end{array}}

\def\G{{\cal G}}
\def\Y{{\cal Y}}

\def\X{{\cal X}}

\def\S{{\cal S}}

\def\L{{\cal L}}
\def\E{{\cal E}}

\def\Z{{\cal Z}}
\def\Q{{\cal Q}}
\def\F{{\cal F}}
\def\I{{\cal I}}
\begin{document}
\begin{center}
{\LARGE {Charging Symmetries and Linearizing Potentials\\
\vskip 0.3cm
for Gravity Models with Symplectic Symmetry
}}
\end{center}
\vskip 1.5cm
\begin{center}
{\bf \large {Oleg Kechkin}}
\end{center}
\begin{center}
Institute of Nuclear Physics,\\
Moscow State University, \\
Moscow 119899, RUSSIA, \\
e-mail: kechkin@monet.npi.msu.su
\end{center}
%%%%%%%%%%%%%%%%%%%%%%%%%%%%%%%%%%%%%%%%%%%%%%%%%%%%%%%%%%%%%%%%%%%%%%%%%%%%%
\vskip 1.5cm
\begin{abstract}
In this paper we continue to study a class of four--dimensional
gravity models with
$n$ Abelian vector fields and $Sp(2n)/U(n)$ coset of scalar fields.
This class contains General Relativity ($n=0$) and Einstein--Maxwell
dilaton--axion theory ($n=1$), which arizes in the low--energy limit
of heterotic string theory. We perform reduction of the model with
arbitrary $n$ to three dimensions and study the subgroup of non--gauge
symmetries of the resulting theory. First, we find an explicit form these
symmetries using Ernst matrix potential formulation. Second, we construct
new matrix variable which linearly transforms under the action of the
non--gauge transformations. Finally, we establish one general invariant
of the non--gauge symmetry subgroup, which allow us to clarify this
subgroup structure.
\end{abstract}
%%%%%%%%%%%%%%%%%%%%%%%%%%%%%%%%%%%%%%%%%%%%%%%%%%%%%%%%%%%%%%%%%%%%%%%%%%%%%%
\newpage
%%%%%%%%%%%%%%%%%%%%%%%%%%%%%%%%%%%%%%%%%%%%%%%%%%%%%%%%%%%%%%%%%%%%%%%%%%
%\section{Introduction}
%%%%%%%%%%%%%%%%%%%%%%%%%%%%%%%%%%%%%%%%%%%%%%%%%%%%%%%%%%%%%%%%%%%%%%%%%%%%%
\setcounter{section}{0}
\setcounter{equation}{0}
\renewcommand{\theequation}{1.\arabic{equation}}
\section*{Introduction}
Four--dimensional gravity models arise in various contexts as some
generalizations of the General Relativity. They differs one of another by
the form of the Lagrangian of matter fields, or by involving
into consideration terms which are non--linear in respect to the
curvature tensor. The most theoretically promissing models are obtained
in the framework of grand unified theories, as it takes place for the models
arising in the low energy limit of (super)string theories compactified
to four dimensions \cite {kir}.

A regular investigation of gravity models is closely related to study of
their symmetries. The most progress in the symmetry analysis was achieved
for the Einstein and Einstein--Maxwell theories (see \cite {rev} for review).
These two theories are the simplest four--dimensional gravity models which
become {\it sygma-models with symmetric target space} after reduction to
three dimensions. The bosonic sector of heterotic string theory leads to
another example of a theory of this class.
The complete list of such theories was established in \cite {bgm}.

In the previous work \cite {ky1} we considered the class of four--dimensional
gravity models which become three--dimensional sigma--models with the target
space possessing a symplectic symmetry. 
This class contains the Einstein--Maxwell theory with dilaton and axion
fields (EMDA), which arises as some truncation of the low--energy heterotic
string effective theory. Moreover, a general representative of
this class is the natural matrix generalization of the EMDA theory. Actually,
the general  {\it symplectic gravity model} (SGM) is described by the
action
\be
^4S = \int d^4x {\mid g \mid}^{\frac {1}{2}} \left \{ - ^4R+
{\rm Tr} \left [ \frac {1}{2}\left (\partial p\,\,p^{-1}\right )^2 -
pFF^T + \frac {1}{3}\left ( pH\right )^2\right ] \right \},
\ee
where $R$ is the Ricci scalar for the metric $g_{\mu \nu}$ of the signature
$+,-,-,-$;
\be
F_{\mu \nu} &=& \partial _{\mu}A_{\nu} -
\partial _{\nu}A_{\mu},
\nonumber\\
H_{\mu \nu \lambda} &=& \partial _{\mu}B_{\nu \lambda}
- \frac {1}{2} (A_{\mu}F^T_{\nu \lambda} + F_{\nu \lambda}A^T_{\mu})
+ {\rm cyclic}.
\nonumber
\ee
Thus, in the special case when all variables are functions one obtains
EMDA in
the Einstein frame. In this simplest case $p= e^{-2\phi}$ has the sense of
string coupling (here $\phi$ is the dilaton), and $B_{\mu \nu}$ is the
antisymmetric
($B_{\mu \nu} = - B_{\nu \mu}$) Kalb--Ramond field. In the general SGM case
$p$ and $B_{\mu \nu}$ obtain two additional indeces
(which are hidden in our notations) and become the symmetric matrices of the
dimension $n$, whereas $A_{\mu}$ becomes the column of the same
dimension. Thus, we consider four--dimensional models with $n$ Abelian vector
fields. The relation of SGM with $n>1$ to any superstring theory is not
clear yet, although symplectic symmetry transformations naturally arise
in the supersymmetry context.

>From the motion equations corresponding to the action (1.1) it follows the
possibility to give the SGM an alternative form on--shell. Actually, using
the pseudoscalar matrix variable $q$, defining by the relation
\be
\nabla _{\sigma}q=\frac{1}{3}E_{\mu\nu\lambda\sigma}pH^{\mu \nu \lambda}p,
\ee
one can rewrite the motion equations in the form which corresponds to
the action
\be
^4S = \int d^4x {\mid g \mid}^{\frac {1}{2}} \left \{ - R +
Tr \left [ \frac {1}{2}\left ( \left ( \nabla p \,\,p^{-1}\right )^2 +
\left ( \nabla q \right )^2\right ) - pFF^T -
q\tilde FF^T\right ] \right \}.
\ee
In \cite {ky1} it was shown that the general SGM allows the $Sp(2n)$ symmetry on
shell. It was established that after the reduction to three dimensions this
symmetry becomes off shell (a Lagrangian) symmetry. Moreover,  
the symmetry group enhanchement takes place: the complete symmetry group of
the resulting
three--dimensional gravity model becomes isomorphic to $Sp(2(n+1))$
(we call it `U--duality' because it appears by the same way as the
three--dimensional U--duality of the effective superstring theories
\cite {ht}).

Below we separate U--duality to the gauge and non--gauge sectors. Next,
we fix the gauge (the trivial field asymptotics) and construct a
representation of the theory which linearizes the non--gauge sector.
Transformations of this sector form a {\it charging symmetry} (CS) subgroup.
They generate charged solutions from neutral ones (see \cite {hk1} for CS
in the heterotic string (HS) theory).

This lettr is organized as follows. In Sec. 2 we review the matrix Ernst
potential (MEP) formulation for SGM reduced to three dimensions \cite {ky1}.
In Sec. 3 we obtain all the CS transformations in a finite form
using the MEP formulation. After that in Sec. 4 we introduce new 
matrix variable and show that this variable transforms {\it linearly} under
the action of the all CS transformations. We derive this {\it linearizing
potential} (LP) for the General Relativity case
using Ernst
potential formulation (the details can be found in the Appendix A), and
directly generalize the result to the general SGM case. After that we
construct one charging symmetry invariant (CSI) which allow us to
establish the CS group structure (some properties of its algebra
are studied in the Appendix B).

We conclude this work with a discussion on the application of CS
transformations to the problem of generation of SGM solutions from the GR
ones.
%%%%%%%%%%%%%%%%%%%%%%%%%%%%%%%%%%%%%%%%%%%%%%%%%%%%%%%%%%%%%%%%%%%%%%%%%%
%\section{Matrix Ernst Potentials}
%%%%%%%%%%%%%%%%%%%%%%%%%%%%%%%%%%%%%%%%%%%%%%%%%%%%%%%%%%%%%%%%%%%%%%%%%%
\setcounter{section}{0}
\setcounter{equation}{0}
\renewcommand{\theequation}{2.\arabic{equation}}
\section*{Matrix Ernst Potential}
%%%%%%%%%%%%%%%%%%%%%%%%%%%%%%%%%%%%%%%%%%%%%%%%%%%%%%%%%%%%%%%%%%%%%%%%%%%
Matrix Ernst potential contains all information about the dynamical
variables of SGM reduced to three dimensions (for definiteness we consider
stationary fields). These variables consist of 

\noindent
a) scalar fields $f=g_{00}$, $v=\sqrt 2A_0$ and $p$;

\noindent
b) pseudoscalar field $\kappa$;

\noindent
c) vector fields $\omega_i=-f^{-1}g_{0i}$ and $A_i$;

\noindent
d) tensor field (three--metric)
$h_{ij}=-fg_{ij}+f^2\omega_i\omega_j$.
In three dimensions, both vector fields $\overrightarrow{A}$ and
$\overrightarrow{\omega}$ can be dualized on--shell:
\be
\nabla\times\overrightarrow{A}&=&\frac{1}{\sqrt 2}\left [ f^{-1}p^{-1}
\left(\nabla u-\kappa\nabla v\right ) + \overrightarrow {\omega}\times
\nabla v\right ],
\nonumber\\
\nabla\times\overrightarrow{\omega}&=&-f^{-2}\left (
\nabla \chi+v^T\nabla u-u^T\nabla v\right ),
\ee
The resulting three--dimensional theory describes the scalars
$f$, $v$ and $p$ and
pseudoscalars $\kappa$, $u$ and $\chi$ coupled to the metric $h_{ij}$.

We define the {\it matrix} \, Ernst potential as follows:
\be
\E=
\left(
\ba{cc}
\E&\F^T\cr
\F&-z
\ea
\right),
\ee
where
\be
z=q+ip,
\quad
\F=u-zv,
\quad
\E=if-\chi+v^T\F.
\ee
Thus, matrix Ernst potential is a complex symmetric $(n+1)\times (n+1)$
matrix. In \cite {ky1} it was shown that all the motion equations can be
derived from the action
\be
^3S&&=
\int d^3xh^{\frac{1}{2}}\left\{
-^3R+{\cal L}_{{}_{SGM}}\right \}
\nonumber\\
&&=
\int d^3xh^{\frac{1}{2}}\left\{
-^3R+2{\rm Tr}\,\left [
\nabla E\,\left ( E-\bar E\right )^{-1}
\nabla \bar E\,\left ( \bar E-E\right )^{-1}\right ].
\right\}
\ee
In the case of $n=0$ our theory becomes the standard General Relativity, and
Eq. (2.4) reproduces a conventional Ernst formulation of the stationary
Einstein
gravity \cite {e}. If $n=1$, one deals with the Einstein-Maxwell theory
with dilaton and axion fields, whose matrix Ernst potential formulation
was proposed in \cite {gk}. We consider these 
theories as two first representatives of the class of gravity models
possessing the symplectic symmetry; all of them allow the
matrix Ernst potential formulation. 
%%%%%%%%%%%%%%%%%%%%%%%%%%%%%%%%%%%%%%%%%%%%%%%%%%%%%%%%%%%%%%%%%%%%%%%%%%
%\section{Charging Symmetries}
\setcounter{section}{0}
\setcounter{equation}{0}
\renewcommand{\theequation}{3.\arabic{equation}}
\section*{Charging Symmetries}
%%%%%%%%%%%%%%%%%%%%%%%%%%%%%%%%%%%%%%%%%%%%%%%%%%%%%%%%%%%%%%%%%%%%%%%%%%%
The complete Lagrangian symmetry group (U--duality) of the 
symplectic gravity model reduced to three dimensions is $Sp(2(n+1))$
\cite {bgm}.
Its action on the matrix Ernst potential
$E$ had been established in \cite {ky1}.
There was shown that the discrete symmetry transformation
\be
\E\rightarrow -\E^{-1}.
\ee
For the case of General Relativity this transformation was established in
\cite {kn}.
In the effective heterotic string theory context (here we have its
truncation for $n=1$) this discrete symmetry is known as strong--weak
coupling duality, or S--duality \cite {kir}. Below we call it
briefly `SWCD'. It is easy to prove that SWCD maps the $E$--shift
symmetry
\be
E\rightarrow E+\lambda
\ee
into the Ehlers transformation
\be
E^{-1}\rightarrow E^{-1}+\epsilon
\ee
where $\lambda$ and $\epsilon$ are the real symmetric matrices
$(\lambda\rightarrow\epsilon)$.
The remaining symmetry is the scaling transformation
\be
E\rightarrow\S^TE\S;
\ee
it is invariant under the action of (3.1) (with
$\S\rightarrow (\S^T)^{-1}$). Thus, the SGM U--duality 
consists of one doublet and one singlet of the
strong--weak coupling duality.

Now let us consider the arbitrary constant potential
$E=E_{{}_{\infty}}$,
which can be interpreted as the asymptotical value of $E$ near to the spatial
infinity.
Applying the shift symmetry with
$\lambda=-{\rm Re}\,(E_{{}_{\infty}})$,
we remove the real part of $E$. Next, the
scaling with
\be
\S=
\left(
\ba{cc}
f^{-\frac{1}{2}}_{{}_{\infty}}&0\cr
-f^{-\frac{1}{2}}_{{}_{\infty}}v_{_{}{\infty}}&\pi^{-1}_{{}_{\infty}}
\ea
\right),
\ee
where $p_{{}_{\infty}}=\pi^T_{{}_{\infty}}
\sigma
\pi_{{}_{\infty}}$,
leads $E$ to its trivial form
\be
\Sigma=
\left(
\ba{cc}
1&0\cr
0&-\sigma
\ea
\right).
\ee
Here
\be
\sigma=
\left(
\ba{cc}
1_{n-k}&0\cr
0&-1_k
\ea
\right)
\nonumber
\ee
is the signature matrix for $p_{{}_{\infty}}$,
whereas
$\pi_{{}_{\infty}}$ is the corresponding tetrad matrix
(for the models with the non--negative energy density $k=0$).
Thus, U--duality
contains gauge
transformations which can be used for the removing of the all field
asymptotics.
Conversely, one can apply these `dressing' transformations to obtain
arbitrary
asymptotics for the originally asymptoticaly--free field
configuration. In the rest part of the letter we fix the gauge and put
$E_{{}_{\infty}}=i\Sigma$.

Transformations preserving fixed asymptotics form a subgroup, which
we call `charging' because these transformations generate charged
solutions from neutral ones.
Scaling transformation contains a part of charging symmetry (CS) subgroup.
Actually, scalings constrained by
\be
\S^T\Sigma\S=\Sigma
\ee
do not change the chosen $E$--asymptotics. Thus, the group of charging
symmetries contains the $SO(n-k,k+1)$ subgroup of the scaling symmetry. We
call the scalings which satisfy Eq. (3.7) {\it normalized scaling
transformation} (NST).

One can see that the Ehlers transformation with the arbitrary non--trivial
parameter $\epsilon$ moves the asymptotical value
$E_{{}_{\infty}}=i\Sigma$.
However, some combination of
the Ehlers transformation with special shift and scaling
duality belongs to the charging symmetry subgroup. Actually, let us
suppose that the Ehlers transformation with the arbitrary antisymmetric
parameter $\epsilon$ is applied to the matrix
$E_{{}_{\infty}}=i\Sigma$. Then
$E_{{}_{\infty}}$ becomes changed. To remove the
real part
of new $E_{{}_{\infty}}$, we perform the shift transformation with
$\lambda=
(i\Sigma -\epsilon)^{-1}\epsilon
(i\Sigma+\epsilon)^{-1}$. After that, we transform the resulting
$E_{{}_{\infty}}$--value to $i\Sigma$ using the scaling (3.4) with 
$\tilde\S$ satisfying the restriction
\be
\tilde\S\Sigma\tilde\S^T=\Sigma+\epsilon\Sigma\epsilon.
\ee
The resulting {\it normalized Ehlers transformation} (NET) has the form
\be
E\rightarrow\tilde\S^T\left [
\left(E^{-1}+\epsilon\right)^{-1}+
(i\Sigma -\epsilon)^{-1}\epsilon
(i\Sigma+\epsilon)^{-1}
\right ]\tilde\S.
\ee

It is easy to see that NST forms the symmetry group of NET itself, because
the condition (3.8) remains unchanged under the action of NST. 

The number of dressing symmetries is equal to the number of
SGM dynamical variables, i.e. to $(n+1)(n+2)$. NST 
gives $(n+1)n/2$ independent parameters. Finally,
NET is defined by the set of $(n+1)(n+2)/2$ parameters
(we fix some $\tilde\S$
satisfying Eq. (3.8)). Thus, all the established 
transformations from the CS subgroup, being independent, are constructed
from $(n+1)^2$ parameters. Then the
common
number of dressing and charging transformations becomes equal to 
$(n+1)(2n+3)$, i.e. to the number
of parameters of the whole U--duality group $Sp(2(n+1))$. From this it
follows that we have found all the gauge (dressing) transformations as
well as all the non--gauge (charging) symmetries. Thus, the CS subgroup
consists of the normalized scaling (3.7) and Ehlers (3.9) transformations.

In the General Relativity case NST is absent (or coinsides with identical
one). Next, NET is related with the single parameter
$\epsilon$; from Eqs. (3.8) and (3.9) it follows that
\be
E\rightarrow \frac {E-\epsilon}{1+\epsilon E},
\ee
Thus, the charging symmetry subgroup of the stationary Einstein gravity
coincides with the one--parametric normalized Ehlers transformation.
%%%%%%%%%%%%%%%%%%%%%%%%%%%%%%%%%%%%%%%%%%%%%%%%%%%%%%%%%%%%%%%%%%%%%%%%%
%\section{Linearizing potential}
%%%%%%%%%%%%%%%%%%%%%%%%%%%%%%%%%%%%%%%%%%%%%%%%%%%%%%%%%%%%%%%%%%%%%%%%%%
\setcounter{section}{0}
\setcounter{equation}{0}
\renewcommand{\theequation}{4.\arabic{equation}}
\section*{Linearizing Potential}
%%%%%%%%%%%%%%%%%%%%%%%%%%%%%%%%%%%%%%%%%%%%%%%%%%%%%%%%%%%%%%%%%%%%%%%%%%%
One can see that the normalized scaling acts as linear
transformation on the matrix Ernst potential $E$, whereas
the normalized Ehlers transformation is some fractional--linear
map. In this section we establish new matrix potential $\Z$
which linearly transforms under the action of the {\it all} CS
transformations, i.e. $\Z$ is a CS linearizing potential.
Our plan is following: we calculate $\Z=\Z (E)$ for the General Relativity
case (n=0) and extend the result to the general SGM case (arbitrary $n$).

In the Appendix A one can find the details of the LP derivation for the
stationary Einstein gravity. The result is:
\be
\Z=2\left(E+i\right)^{-1}+i.
\ee
Thus, $\Z_{{}_{\infty}}=\Z (E_{{}_{\infty}})=\Z (i)=0$, i.e. the near to
spatial infinity asymptotics are trivial.
The relation (4.1) admits a straightforward generalization
to the case of matrix variables. Actually, the simple substitution
$i\rightarrow i\Sigma$ preserves triviality of $\Z_{{}_{\infty}}$; using
it, we obtain:
\be
\Z=2\left(E+i\Sigma\right)^{-1}+i\Sigma.
\ee

To verify that the fractional--linear function (4.2) defines the
SGM linearizing potential, one must rewrite
all the CS transformations in terms of the $\Z$--representation.
For NST one immediately obtains:
\be
\Z\rightarrow\S^{-1}\Z(\tilde\S^T)^{-1}.
\qquad \left( {\rm NST
}\right)
\ee
After some amount of algebra based on the use of the relation (3.8), one
establishes that NET also has a linear form:
\be
\Z\rightarrow \tilde\S^{-1}\left ( 1-i\epsilon\Sigma\right )
\Z\left ( 1-i\Sigma\epsilon\right ) \left ( \tilde\S^T\right )^{-1}.
\qquad \left( {\rm NET
}\right).
\ee
Thus, the introduced
matrix $\Z$ actually is a linearizing potential of the
charging symmetry subgroup of the stationary symplectic gravity
model with arbitrary $n$.

To analyse the CS group structure we will need in one general CS invariant.
This invariant can be `extracted' from the Lagrangian $\L_{{}_{SGM}}$
(see Eq. (2.4)). To do this, let us consider asymptotically trivial fields
with the non--zero Coulomb terms: 
\be
\Z=\frac{\Q}{r}+o\left(\frac{1}{r}\right),
\ee
where $\Q$ is a charge matrix and $r$ tends to the spatial
infinity. Then, from Eq. (4.2) we obtain that
\be
\L_{{}_{SGM}}=\frac{1}{2r^4}
{\rm Tr}\left \{ \bar\Q\Sigma\Q\Sigma\right \}
+o\left(\frac{1}{r^4}\right).
\ee
The quadratic charge combination
\be
\I(\Q)=
{\rm Tr}\left \{ \bar\Q\Sigma\Q\Sigma\right \}
\ee
is a CS invariant,
because $\L_{{}_{SGM}}$ is the CS invariant and all
its
terms related to the $1/r$ power expansion are also CS invariants.
Then, from Eq. (4.5) it follows that the charge and
linearizing potential matrices have the same transformation properties. Thus,
the function
\be
\I(\Z)={\rm Tr}\left \{\bar\Z\Sigma\Z\,\Sigma\right \}
\ee
must be a charging symmetry invariant (see \cite {emda} for EMDA).

One can see that the charging symmetry transformations are of the form
\be
\Z\rightarrow\G^T_i\Z\G_i,
\ee
where $i=$ NST and NET. 
An explicit
form of the matrices $\G_{{}_{NST}}$ and $\G_{{}_{NET}}$
can be obtained from Eqs. (4.3) and (4.4). These are:
\be
\G_{{}_{NST}}=\left ( \S^T\right )^{-1},\quad
\G_{{}_{NET}}=\left ( 1-i\Sigma\epsilon\right )
\left ( \tilde\S^T\right )^{-1},
\ee
where $\S$ and $\tilde\S$ satisfy Eqs. (3.7) and (3.8) correspondingly.

To preserve $\I(\Z)$ both transformations must satisfy the
$U(n-k,k+1)$ group relation
\be
\G^+_i\,\Sigma\,\G_i\,=\,\Sigma.
\ee
This really takes place; thus
$\G_i\in U(n-k,k+1)$. Now let us note that the common number of independent
parameters of
NST and NET is $(n+1)^2$, i.e. the same one
as for the group $U(n-k,k+1)$. Moreover, if we consider the infinitesimal
transformations $\Gamma_i$ ($\G_i=e^{\Gamma_i}$ and
compute $\Gamma =\sum_i \Gamma_i$, we obtain
\be
\Gamma=-i\Sigma\epsilon-\sigma^T,
\ee
where $\sigma$ denotes the NST generator ($\S=e^{\sigma}$).
This matrix is a {\it general} solution of the equation
$\left (\Gamma\right )^+=\,-\,\Sigma\,\Gamma^{{}^{right}}\,\Sigma$,
which defines the $u(n-k,k+1)$ algebra (its structure is discussed
in the Appendix B).
>From this we conclude that the general CS transformation
matrix is the general matrix of the group
$U(n-k,k+1)$. It can be constructed as the product of NST and NET
matrices multiplied in an arbitrary order.

Thus, we have established the following {\it simplest} form of the charging
symmetry transformations:
\be
\Z\rightarrow\G^T\Z\,\G,
\quad {\rm where} \quad \G\in U(n-k,k+1).
\ee
It is important to note that the transformation of the charge matrix
$\Q$ can be obtained from Eq. (4.13) using the replacement
$\Z\rightarrow\Q$.

The strong--weak
coupling duality transformation in terms of the linearizing potential
$\Z$ takes the form:
\be
\Z\rightarrow -\Sigma\,\Z\,\Sigma.
\ee
>From Eqs. (3.1) and (4.2) it follows that SWCD acts on
$\G$ in the following way:
\be
\G\rightarrow\Sigma\G\Sigma
\ee
One can see that this map preserves the group relation (4.11).
This means, that the whole CS subgroup is also SWCD invariant.
Taking into
account that the dressing symmetries do not possess this property we
obtain the following alternative definition of the CS subgroup:
{\it the charging
symmetry subgroup is the maximal subgroup of the U--duality which is
invariant
under the action of the SWCD transformation}.
%%%%%%%%%%%%%%%%%%%%%%%%%%%%%%%%%%%%%%%%%%%%%%%%%%%%%%%%%%%%%%%%%%%%
%\section{Concluding Remarks}
%%%%%%%%%%%%%%%%%%%%%%%%%%%%%%%%%%%%%%%%%%%%%%%%%%%%%%%%%%%%%%%%%%%%%%%%%%
\setcounter{section}{0}
\setcounter{equation}{0}
\renewcommand{\theequation}{7.\arabic{equation}}
\section*{Concluding Remarks}
%%%%%%%%%%%%%%%%%%%%%%%%%%%%%%%%%%%%%%%%%%%%%%%%%%%%%%%%%%%%%%%%%%%%%%%%%%%
Thus, we have extracted all the charging symmetry transformations
from the general Lagrangian symmetry
group of the general
four--dimensional symplectic gravity model reduced to three dimensions.
We have established matrix linearizing potential which undergo
{\it linear
homogeneous transformations} when the charging symmetries act. We have
constructed one general invariant of these symmetries, quadratic on the
linearizing potentials, and studied the charging symmetry group
structure.

Found representation can be applied to the problem of generation of SGM
solutions from the GR ones. It is abvious that the LP formalism is the
most convenient for the generation of solutions trivial at the spatial
infinity (in the three--dimensional sense, see
Eq. (4.5); in four dimensions these solutions allow, for example, the
NUT charge). Actually,
starting from the arbitrary GR solution, rewritten in the LP form, and
applying the SGM charging symmetry transformations accordingly Eq. (4.13),
one obtains a class of SGM solutions with the manifest $U(n-k,k+1)$
symmetry.
Next, the LP formulation gives the most natural tool for the
construction of the extremal
(with $h_{ij}=\delta_{ij}$) Israel--Wilson--Perj'es--like solutions
\cite {iwp}; these solutions form the CS--invariant class.
There are some
other directions in applying of the LP formulation; all of them are
based on the fact of linearization of the non--gauge symmetries of the
stationary symplectic gravity models.
%%%%%%%%%%%%%%%%%%%%%%%%%%%%%%%%%%%%%%%%%%%%%%%%%%%%%%%%%%%%%%%%%%%%%%%%%%%
\section*{Acknowledgments}
I thank my colleagues for encouraging. 
%%%%%%%%%%%%%%%%%%%%%%%%%%%%%%%%%%%%%%%%%%%%%%%%%%%%%%%%%%%%%%%%%%%%%%%%%%%%
\setcounter{section}{0}
\setcounter{equation}{0}
\renewcommand{\theequation}{A.\arabic{equation}}
\section*{Appendix A: LP for GR}
In this Appendix we derive linearizing potential of the
charging symmetry subgroup for the stationary General Relativity.

The generator of the CS subgroup can be obtained from Eq. (3.10);
the result is:
\be
\X =
-\left ( E^2+1\right ) \partial_E
\ee
(we write down only the holomorphic part of generators).

The CS transformation (NET) can be realized $linearly$ in the following way.
Let $\Z$ be the complex variable whose finite
transformation has a transparent $U(1)$ form:
\be
\Z\rightarrow e^{2i\alpha}\Z,
\ee
where $\alpha$ is a real parameter.
The corresponding generator is:
\be
\Y=2i\Z\partial_{\Z}.
\ee

Now we identify the generators $\X$ and $\Y$; they are equal
up to a real constant factor:
\be
\X=c\Y.
\ee
Supposing that the functional relation
$E=E(\Z)$ exists, we obtain the differential
equation $2c\Z E_{,\Z}=i(E^2+1)$ of the first order which defines it. 
Solving this equation, we obtain:
\be
\Z=c'\left ( \frac {E+i}{E-i}\right )^c,
\ee
where $c'$ is an arbitrary complex constant. We choose $c=-1$ in order to
reach the simplest possible fractional--linear form of
the constructed solution (it is important for the matrix generalization of
Eq. (A.5)) and the relation $\Z(i)=0$. Thus, we choose LP to be trivial at the
spatial infinity. Next, the concrete value of $c'$ is not important, and
we put $c'=i$ for the simplest form of the result. Finally, the relation
between the linearizing and Ernst potential takes the form:
\be
\Z=\frac {2}{E+i}+i.
\ee
It is easy to see that if $\Z$ linearizes some 
transformation, then $c'\Z^{-c}$ also will be a linearizing potential. This
explains appearing of two arbitrary constants in Eq. (A.5). 
%%%%%%%%%%%%%%%%%%%%%%%%%%%%%%%%%%%%%%%%%%%%%%%%%%%%%%%%%%%%%%%%%%%%%%%%%%%%%
\setcounter{section}{0}
\setcounter{equation}{0}
\renewcommand{\theequation}{B.\arabic{equation}}
\section*{Appendix B: CS Algebra for SGM}
In this Appendix we compute the commutation relations for the charging
symmetry algebra of the symplectic gravity model with arbitrary $n$.

The generators were constructed in the infinitesimal form; this means
that $\epsilon=\xi e$ and $\sigma=\xi s$, where
$\xi$ is the infinithesimal parameter. Here the matrices $e$ and $s$ 
are finite ($e^T=e$, $s^T=-\Sigma s\Sigma$); they define the finite form
of the NST and NET
generators:
\be
\Gamma_{{}_{NST}}\left ( s\right )=-s^T,\quad
\Gamma_{{}_{NET}}\left ( e\right )=-i\Sigma e.
\ee
The computation of the commutation relations gives:
\be
&&\left [ \Gamma_{{}_{NST}}\left ( s'\right ),
\Gamma_{{}_{NST}}\left ( s^{''}\right )\right ]
=\Gamma_{{}_{NST}}\left ( \left [
s',s^{''}\right ]\right ),
\nonumber\\
&&\left [ \Gamma_{{}_{NET}}\left ( e'\right ),
\Gamma_{{}_{NET}}\left ( e^{''}\right )\right ]
=-\Gamma_{{}_{NST}}\left ( \left [
\Sigma e',\Sigma e^{''}\right ]\right ),
\nonumber\\
&&\left [ \Gamma_{{}_{NET}}\left ( e\right ),
\Gamma_{{}_{NST}}\left ( s\right )\right ]
=-\Gamma_{{}_{NET}}\left ( se+es^T\right ),
\ee
One can see, that only NST generators form a subalgebra, and the 
minimal algebra including NET is equal to the full CS algebra.
%%%%%%%%%%%%%%%%%%%%%%%%%%%%%%%%%%%%%%%%%%%%%%%%%%%%%%%%%%%%%%%%%%%%%%%%%%%%%%%
%\newpage


\begin{thebibliography}{30}
\bibitem{kir}
E. Kiritsis,
{\it Introduction to Superstring Theory}, CERN--TH/97-218, 
hep--th/9709062.
\bibitem{rev}
D. Kramer, H. Stephani, M. McCallum, E. Herlt,
{\it Exact Solutions of the Einstein Field Equations}, Deutcher Verlag
der Wissenschaften, Berlin, 1980.
\bibitem{bgm}
P. Breitenlohner, G. Gibbons and D. Maison,
Commun. Math. Phys. {\bf 120} (1987) 295;

\bibitem{ky1}
O. Kechkin and M. Yurova,
J. Math. Phys. {\bf 39} (1998) 5446.
\bibitem{ht}
C.M. Hull and P.K. Townsend,
Nucl. Phys. {\bf B438} (1995) 109. 
\bibitem{hk1}
A. Herrera-Aguilar and O. Kechkin,
{\it Charging Symmetries and Linearizing Potentials for Heterotic
String in Three Dimensions},
hep-th/9811189.
\bibitem{kn}
D. Kramer, G. Neugebauer and H. Stephani,
Fortschr. Physik, {\bf 24}, 59.
\bibitem{gk}
D.V. Gal'tsov and O.V. Kechkin, Phys. Lett. {\bf B361} (1995) 52.
\bibitem{e}
F.J. Ernst, Phys. Rev. {\bf 167} 1175, (1968) 
\bibitem{emda}
O. Kechkin and M. Yurova,
Gen. Rel. Grav. {\bf 29}, 10, (1997) 1283;
A. Herrera-Aguilar, O. Kechkin
Mod. Phys. Lett. A13: 1907-1914, 1998
\bibitem{iwp}
W. Israel and G.A. Wilson,
J. Math. Phys. {\bf 13} (1972) 865;
Z. Perj\'es, Phys. Rev. Lett. {\bf 27} (1971) 1668.
\end{thebibliography}
\end{document}